\documentclass[a4paper,11pt]{article}
\usepackage{a4wide}
\usepackage[dvipdfmx]{graphicx}
\usepackage{amsmath,amssymb}
\usepackage{bm}
\usepackage{subcaption}
\usepackage{color}
\usepackage{cite}

\usepackage{unit}

\title{Control of Oscillatory Temperature Field in a Building\\
via Damping Assignment to Nonlinear Koopman Mode\thanks{This research was supported in part by JST, PRESTO Grant No. JP-MJPR1926.}
}

\author{
Yoshihiko Susuki, Kohei Eto, Naoto Hiramatsu, and Atsushi Ishigame\thanks{They were with Department of Electrical and Information Systems, Osaka Prefecture University, 
        1-1 Gakuen-cho, Naka-ku, Sakai 599-8531, Japan. 
        Presently, Y. Susuki is with Department of Electrical Engineering, Kyoto University, 
        Katsura, Nishikyo-ku, Kyoto 615-8510, Japan.  
        {\tt susuki.yoshihiko.5c@kyoto-u.ac.jp}}
}
\date{}
\begin{document}
\maketitle

\begin{abstract}
This paper addresses a control problem on air-conditioning systems in buildings that is regarded as a control practice of nonlinear distributed-parameter systems. 
Specifically, we consider the design of a controller for suppressing an oscillatory response of in-room temperature field. 
The main idea in this paper is to apply the emergent theory of Koopman operator and Koopman mode decomposition for nonlinear systems, and to formulate a technique of damping assignment to a nonlinear Koopman mode in a fully data-driven manner. 
Its effectiveness is examined by numerical simulations guided by measurement of a practical room space.  
\end{abstract}

\section{Introduction}

Building thermal control has attracted a lot of interest in not only building physics and science but also control engineering: see, e.g., \cite{Wen:2018,Drgona_ARC50}.  
The target dynamics are complicated physical phenomena due to interactions of energy transport, fluid motion, air conditioning control, and human movement, which are typically modeled by a form of nonlinear Partial Differential Equations (PDEs). 
Recent attention to data-driven science and artificial intelligence has opened a new direction of the building thermal control, in which the utility of measurement data and computation will be maximized: see, e.g., \cite{Drgona_EB243,Nagy:2021}. 
It is thus of technological significance to develop a data-driven methodology for handling such nonlinear multi-scale dynamics for the next-generation building thermal control. 

Recently, the Koopman operator framework, especially Koopman Mode Decomposition (KMD), has been applied to data analysis of complicated thermal dynamics in buildings: see, e.g., \cite{Eisenhower_SimBuild2010,Georgescu_SimBulid2012,Georgescu_EB86,Kono_TSICE53a,Masaki_IOP238,Kono-Ch18:2020,Hiramatsu_PRE102,Boskic_Energies14}. 
KMD is a technique of nonlinear time-series analysis that decomposes multivariate time series into modes oscillating at a single frequency \cite{Mezic_ND41,Rowley_JFM641}. 
The technique is mathematically guided by Koopman operator theory of nonlinear dynamical systems (see, e.g., \cite{Marko_CHAOS22,Mezic_ARFM45}) that are regarded as mathematical models of the target time series.  
In KMD, the frequency and damping rate of each mode are determined by an eigenvalue of the Koopman operator, called Koopman eigenvalue, 
and the mode vector is by the associated eigenfunction of the Koopman operator, called Koopman eigenfunction.  
Significant abilities of KMD are to handle dynamics arising due to nonlinearity of the underlying model and to clearly separate multiple time scales embedded in the time series \cite{Susuki_IEEETPWRS26}, which are suitable to the analysis of building thermal dynamics. 

In this paper, we report our research on how the Koopman operator framework is utilized for control of thermal dynamics in buildings. 
The framework itself has been applied to control theory and practice: see, e.g., \cite{Brunton:2019,Abraham_IEEETR35,TheBook:2020,Otto_ARCR4,Bruder_IEEETR37,Bevanda_Preprint:2021,Harrison_Mathematics9}.
In \cite{Hiramatsu_PRE102} we reported the KMD of measured data on oscillatory responses of temperature field inside a room space. 
The oscillations are mainly caused by the VAV (Variable Air Volume) operation of commercial Packaged-Type Air Conditioners (PTACs) (see, e.g., \cite{Kono_JBPS11}), and their suppression has an technological benefit for reducing energy use and enhancing human comfort. 
In this paper, we propose the design of a controller for suppressing such an oscillatory response of the in-room temperature field by utilizing spectral properties of the Koopman operator. 
Technically, we derive a state-feedback law based on the Koopman eigenfunction for a latent nonlinear-system model that enables \emph{damping assignment to a Koopman mode} representing the target oscillatory response.   
The proposed damping assignment is fully data-driven in the sense that the Koopman eigenfunction is estimated directly from measured data on the temperature field without any development of the latent model. 
In this paper, we derive the state-feedback law and present a series of numerical simulations of the target thermal dynamics with and without control. 
Its performance evaluation is presented in terms of level sets of the Koopman eigenfunction. 
These numerical simulations are guided by measurement of a practical room space reported in \cite{Hiramatsu_PRE102}. 
This paper is a substantially extended version of non-reviewed proceedings \cite{Eto_SCI22,Eto_JJACC21}. 

The rest of this paper is organized as follows. 
Sec.\,\ref{sec:2} describes the room space to be controlled and measurement data on the target oscillatory response. 
Sec.\,\ref{sec:3} summarizes the theory of Koopman operators for nonlinear systems and KMD. 
Sec.\,\ref{sec:4} shows the main idea of this control paper: data-driven damping assignment to a nonlinear Koopman mode. 
Sec.\,\ref{sec:5} presents a set of numerical simulations to demonstrate the main idea. 
Conclusions of this paper are made in Sec.\,\ref{sec:6}. 

\section{Controlled Object}
\label{sec:2}

\begin{figure}[t]
\centering
\includegraphics[width=0.45\textwidth]{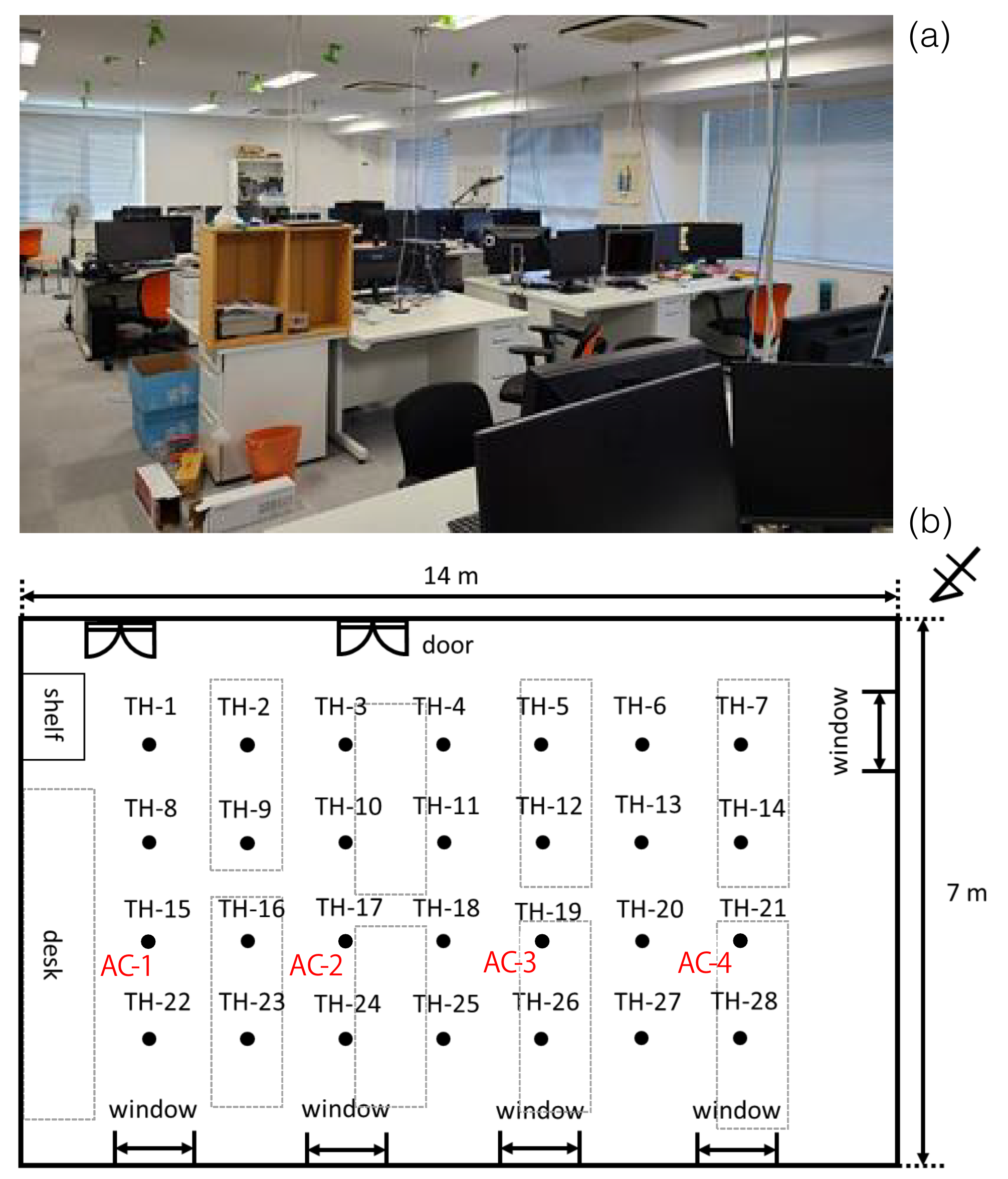}
\caption{Photograph and geometry of room space studied in this paper.}
\label{fig1.5}
\vspace*{2mm}
\includegraphics[width=0.48\textwidth]{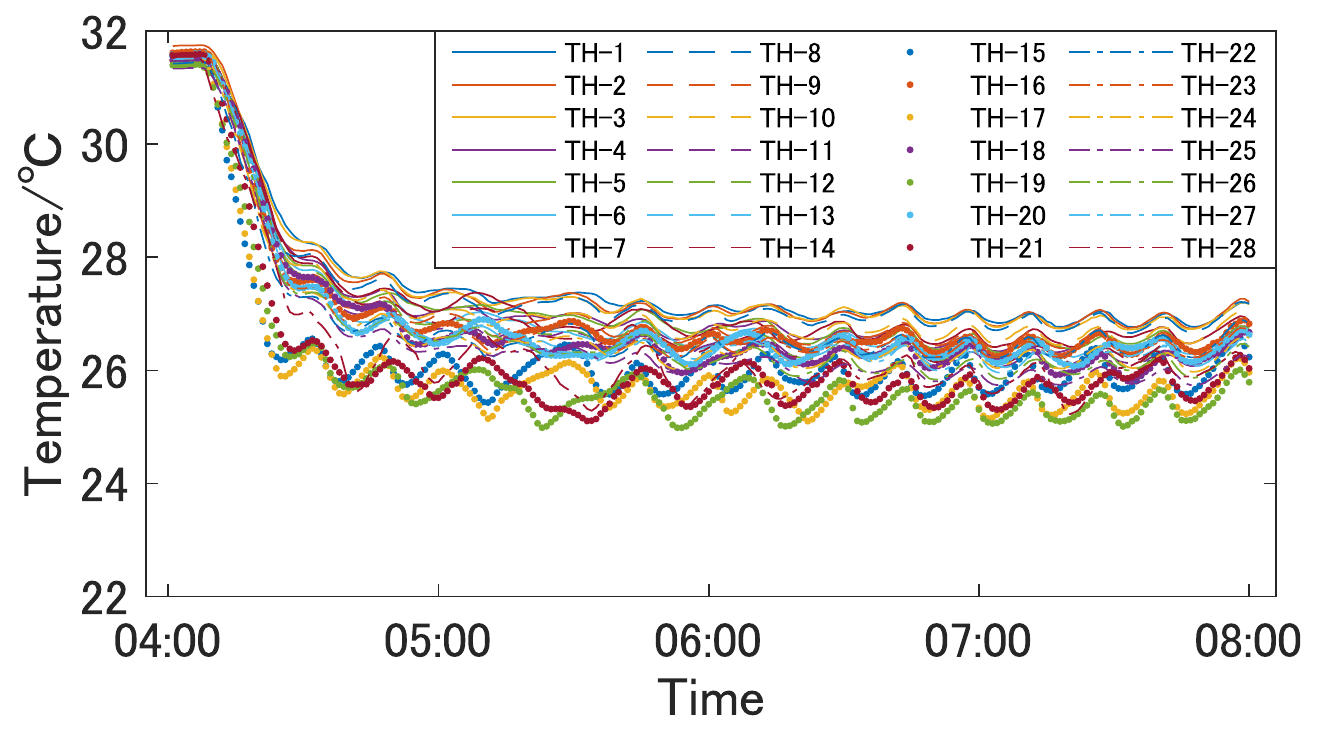} 
\caption{Measurement data of oscillatory temperature field in the room space.  
The set-point of temperature in PTACs is 27\,deg.C.}
\label{fig:measurement}
\end{figure}

\subsection{Room Space and Measurement Data}

In this section, we describe the room space to be controlled, which is a research room located in Nakamozu Campus, Osaka Prefecture University, Sakai, Japan. 
Fig.\,\ref{fig1.5} shows a photograph of the room space and its geometrical overview from the top (ceiling) of the space. 
The width of the room is about 14\,m, the depth is about 7\,m, and the height is about 2.6\,m. 
The four PTACs (Package-Type Air Conditioners), denoted by \textsf{AC-1} to \textsf{AC-4}, are installed on the ceiling and supply air to their neighborhoods.  
In Fig. \ref{fig1.5}, \textsf{TH-1} to \textsf{TH-28} indicate the locations for measurement of in-room temperature used in \cite{Hiramatsu_PRE102}. 
The temperature data measured around the four PTACs, that is, at the four locations, 
\textsf{TH-15}, \textsf{17}, \textsf{19}, and \textsf{21}, will be addressed for the control design in Sec.\,\ref{sec:4}. 

Figure~\ref{fig:measurement} shows the measurement data on temperature field in the room space. 
The measurement was conducted from 4am to 8pm in summer under the cooling operation of PTACs. 
It is observed in the figure that the temperature measured at the total 28 locations oscillate in time. 
The oscillation arises around the set-point of PTACs, 27\,deg.C, and its period is roughly {\color{black}15 to 20\,min}. 
The oscillation is the object to be suppressed in this paper. 

\subsection{Mathematical Model for Simulation Studies}
\label{subsec:infinite-model}

We next introduce a mathematical model for temperature field used for simulation studies. 
Following \cite{Hiramatsu_PRE102}, the temperature field is represented by the scalar field in two dimension (2d), denoted by $\theta(\vct{r},t)$ with the location $\vct{r}=(x,y)^\top\in D\subset\mathbb{R}^2$ and time $t\in\mathbb{R}$. 
The $x$ direction is for the width (horizontal) in Fig.\,\ref{fig1.5} and the $y$ direction for the depth (vertical). 
The symbol $\top$ stands for the transpose of vectors.  
The 2d problem formulation is relevant to clarifying the thermal dynamics of the same room space as shown in \cite{Hiramatsu_PRE102}. 
Based on \cite{Kono_JBPS11}, we use the so-called \emph{effective diffusion} model (linear PDE) for the field's dynamics as follows: 
\begin{equation}
\frac{\DD}{\DD  t}\theta(\vct{r},t)=D\sub{eff}\!\left(\frac{\DD^2}{\DD x^2}+\frac{\DD^2}{\DD y^2}\right)
\theta(\vct{r},t)+\frac{p\sub{air}(\vct{r},t)}{\rho c\sub{p}}.
\label{eqn:拡散方程式}
\end{equation} 
The parameter $D\sub{eff}$ is the effective diffusion assumed to be constant in location and time, $\rho$ the density of air, and $c_{\rm}$ its specific heat at  constant pressure. 
The variable $p\sub{air}$ is the heat input per unit time and volume from the four PTACs and is represented via the method of bulk convection \cite{Hensen:2011} as follows: 
\begin{equation}
p\sub{air}(\vct{r}\sub{h},t)=\frac{v\sub{air}(\vct{r}\sub{h},t)\{\theta\sub{air}(\vct{r}\sub{h},t)-\theta\sub{p}(\vct{r}\sub{h},t)\}}
{V_0(\vct{r}\sub{h},t)}{\rho c\sub{p}},
\label{eqn:BulkConvection}
\end{equation}
where $\vct{r}\sub{h}$ stands for the location of each of the four PTACs (\textsf{AC-1} to \textsf{AC-4}), $v\sub{air}$ for the outlet volume per unit time for each of them, $\theta\sub{air}$ for the outlet temperature to be constant in this paper, and $V_0$ for {\color{black}the unit volume centered at the outlet}.  
The temperature $\theta\sub{p}(\vct{r}\sub{h},t)$ is the room temperature affected by a dynamic effect in the PTAC and is represented by the following first-lag system with delay:
\begin{equation}
\frac{\dd}{\dd t}\theta\sub{p}(\vct{r}\sub{h},t)
=-\frac{1}{T\sub{s}}\{\theta\sub{p}(\vct{r}\sub{h},t)-G\theta(\vct{r}\sub{h},t-L)\},
\label{eqn:PTAC}
\end{equation}
where $T\sub{s}$ is the time constant, $G$ the gain constant, and $L$ the delay constant. 
To maintain the temperature at the cooling situation, we suppose that the outlet volume is regulated according to the switching law, given by
\begin{equation}
v\sub{air}(\vct{r}\sub{h},t)=\left\{
\begin{array}{ll}
V^{\rm on}\sub{air} & (\theta(\vct{r}\sub{h},t-L)>\theta\sub{ref}),\\\noalign{\vskip 1mm}
V^{\rm off}\sub{air} & ({\rm otherwise}),
\end{array}
\right.
\label{eqn:VAV}
\end{equation}
where $\theta\sub{ref}$ is the set-point value, and $V^{\rm on}\sub{air}$ and  $V^{\rm off}\sub{air}$ the values of outlet volume fixed by design, satisfying $V^{\rm on}\sub{air}>V^{\rm off}\sub{air}$. 
The switching law \eqref{eqn:VAV}, known as the VAV (Variable Air Volume) operation in the PTAC, is one of major causes of the oscillatory response in Fig.\,\ref{fig:measurement}.   


\section{Summarized Theory of Koopman Operators}
\label{sec:3}

In this section, we briefly introduce the Koopman operator, Koopman eigenvalues, and Koopman eigenfunctions 
based on \cite{Mezic_ARFM45,TheBook:2020}. 
For this, consider a continuous-time dynamical system described by the following nonlinear Ordinary Differential Equation (ODE):
\begin{equation}
\frac{\dd\vct{x}}{\dd t}=\vct{\dot{x}}=\vct{F}(\vct{x}),
\qquad \vct{x}\in\mathbb{R}^n,
\label{1}
\end{equation}
where $\vct{x}$ is the state variable, and $\vct{F}:\mathbb{R}^n\rightarrow\mathbb{R}^n$ is a nonlinear function (vector-field). 
By assuming that there exists a unique solution of \eqref{1}, 
the \emph{flow} $\vct{S}^t:\mathbb{R}^n\rightarrow\mathbb{R}^n$ is defined as a one-parameter group of nonlinear maps with $t\in\mathbb{R}$, satisfying
\begin{equation}
\frac{\dd}{\dd t}\vct{S}^t(\vct{x})=\vct{F}(\vct{S}^t(\vct{x})).
\label{eqn:流れとODE}
\end{equation}
Here, let us introduce a scalar-valued continuous function defined on the state space, $f: \mathbb{R}^n\rightarrow\mathbb{C}$, which we call the \emph{observable}. 
A linear space of such observables is denoted by $\mathcal{K}$. 
Then, the linear operator ${U}^t:\mathcal{K}\to\mathcal{K}$ is defined as the composition operation of $f$ with $\vct{S}^t$: 
\begin{equation}
{U}^tf:=f\circ\vct{S}^t.
\label{2}
\end{equation}
This ${U}^t$ is called the \emph{Koopman operator} and represents the time evolution of the observable $f$ under the flow $\vct{S}^t$. 
The linearity of ${U}^t$ can be utilized for analysis and design of systems with nonlinear dynamics, which is termed as the Koopman operator framework \cite{TheBook:2020}. 

A significant outcome of the linearity is to introduce the notion of spectra for the nonlinear system \eqref{1}. 
In particular,  the \emph{Koopman eigenvalue} $\nu\in\mathbb{C}$ of ${U}^t$ and associated \emph{Koopman eigenfunction} $\phi_\nu\in\mathcal{K}\setminus\{0\}$ are introduced as follows:
\begin{equation}
{U}^t\phi_\nu={\rm e}^{\nu t}\phi_\nu.
\label{3}
\end{equation}
By chain rule of differentiation, we formally have
\begin{align}
\left.\frac{\dd}{\dd t}({U}^t\phi_\nu)(\vct{x})\right|_{t=0} &= \left.\frac{\dd}{\dd t}\ee^{\nu t}\phi_\nu(\vct{x})\right|_{t=0} \nonumber\\
\{\vct{\nabla}\phi_\nu(\vct{x})\}^\top\left.\frac{\dd}{\dd t}\vct{S}^t(\vct{x})\right|_{t=0} 
&= \left.\frac{\dd}{\dd t}\ee^{\nu t}\right|_{t=0}\phi_\nu(\vct{x}) \nonumber\\
\vct{\nabla}\phi_\nu(\vct{x})^\top\vct{F}(\vct{x}) 
&= \nu\phi_\nu(\vct{x})
\quad (\because \textrm{\eqref{eqn:流れとODE}}),
\label{4}
\end{align}
where $\vct{\nabla}$ is the gradient operator in $\mathbb{R}^n$. 
The Koopman eigenfunction is a key enabler for the control design in this paper. 
Its numerical estimation is proposed in \cite{Matt_JNLS25,Klus_JCD3}, which is termed as the Extended Dynamic Mode Decomposition (EDMD). 
In EDMD, a set of user-specified basis functions (observables) with $q$ number,  $\vct{\gamma}(\vct{x}):=[\gamma_{1}(\vct{x}),\ldots,\gamma_{q}(\vct{x})]^{\top}$: $\mathbb{R}^{n}\rightarrow\mathbb{R}^{q}\ 
(\gamma_{j}\in\mathcal{K})$ are chosen. 
The Koopman eigenfunction $\phi_\nu(\vct{x})$ is then approximated as their linear combination, given by
\begin{equation}
\phi_\nu(\vct{x})\sim\vct{\xi}^{\top}_{\nu}\vct{\gamma}(\vct{x}),
\label{estimated}
\end{equation}
where $\vct{\xi}_\nu\in\mathbb{C}^q$ is a constant vector estimated from time series data of the state's dynamics sampled from the trajectory $\vct{x}(t)=\vct{S}^t(\vct{x}_0)$ ($\vct{x}_0$ is an initial state): see Appendix~\ref{app:EDMD} for details. 

A successful use of the Koopman eigenvalue and eigenfunction is the KMD (Koopman Mode Decomposition). 
Following \cite{Mezic_ARFM45}, under certain conditions of the system \eqref{1} and the space $\mathcal{K}$ of $f$, the state $\vct{x}$ can be expanded into a (countable) set of Koopman eigenfunctions as
\begin{equation}
\vct{x}=\sum^\infty_{j=1}\phi_{\nu_j}(\vct{x})\vct{V}_j,
\label{eqn:KMD0}
\end{equation} 
where $\phi_{\nu_j}\in\mathcal{K}\setminus\{0\}$ is the $j$-th Koopman eigenfunction. 
The constant vector $\vct{V}_j\in\mathbb{C}^n$ is called the \emph{Koopman mode} as a vector-valued coefficient for the expansion.
By acting ${U}^t$ to both sides of \eqref{eqn:KMD0} and using its linearity and eigen-properties, the state trajectory $\vct{x}(t)=\vct{S}^t(\vct{x}_0)$ starting from $\vct{x}_0$ is expanded as
\begin{equation}
\vct{x}(t) ={U}^t\!\!\left.\left(\sum^\infty_{j=1}\phi_{\nu_j}(\vct{x})\vct{V}_j\right)\right|_{x=x_0}
=\sum^\infty_{j=1}\ee^{\nu_j t}\phi_{\nu_j}(\vct{x}_0)\vct{V}_j,
\label{eqn:KMD}
\end{equation}
where $\nu_j\in\mathbb{C}$ is the $j$-th Koopman eigenvalue to which $\phi_{\nu_j}$ belongs. 
If $\nu_j$ is a pure imaginary number as $\ii\omega_j$, where $\ii$ is the imaginary unit and $\omega_j\in\mathbb{R}$,  then the corresponding response $\ee^{\ii\omega_j t}\phi_{\ii\omega_j}(\vct{x}_0)\vct{V}_j$ represents a sustained single-frequency component embedded in $\vct{x}(t)$.
This observation is the starting point of our control design in Sec.\,\ref{subsec:導出}. 

\section{Proposed Design: Damping Assignment to Koopman Mode}
\label{sec:4}

This section provides the main idea of the control design: a technique of damping assignment to a Koopman mode using the Koopman eigenfunction. 

\subsection{Finite-Dimensional Reduction}

The model in Sec.\,\ref{subsec:infinite-model}, 
which is the linear PDE \eqref{eqn:拡散方程式} with the discontinuous nonlinearity \eqref{eqn:VAV}, is a nonlinear infinite-dimensional system and so not easily utilized for the control design. 
In this paper, motivated by full use of temperature data measured in the target space, we suppose that the measured data are represented by a control-affine, finite-dimensional dynamical system as follows: 
\begin{equation}
\vct{\dot{x}}=\vct{F}(\vct{x})+\mathsf{B}\vct{u},\qquad \vct{x}\in\mathbb{R}^M ,\ \vct{u}\in\mathbb{R}^m.
\label{13}
\end{equation}
The state variable $\vct{x}$ in \eqref{13} corresponds to the oscillatory component embedded in the temperature measured at the $M=4$ measurement points \textsf{TH-15} (for \textsf{AC-1}), \textsf{TH-17} (for \textsf{AC-2}), \textsf{TH-19} (for \textsf{AC-3}), and \textsf{TH-21} (for \textsf{AC-4}): $\vct{x}=[x_1,x_2,x_3,x_4]^\top$. 
The drift part in \eqref{13}, namely, $\vct{F}(\vct{x})$ is a model that represents the oscillatory response of the measured temperature. 
In addition, $\mathsf{B}$ is a constant matrix of $M\times m$ dimensions.  
The input $\vct{u}$ is regarded as the heat input from ancillary ACs, and their number $m$ is supposed to coincide with the number $M$ of the existing PTACs. 
In this setting, the heat input $p\sub{air}$ in \eqref{eqn:BulkConvection} is modified with the ancillary input $\vct{u}=[u_1,u_2,u_3,u_4]^\top$ as 
\begin{equation}
p\sub{air}(\vct{r}\sub{h}(\textsf{AC-}i),t)+u_i(t),
\end{equation}
where $\vct{r}\sub{h}(\textsf{AC-}i)$ denotes the location of the $i$-th PTAC for $i=1,\ldots,4$.  
Note that the finite-dimensional system \eqref{13} is {\color{black}supposed} 
to exist implicitly (i.e. latent), but its equation is not used in the control design. 
In Sec.\,\ref{subsec:results}, the Koopman eigenfunction for the drift part of \eqref{13} will be estimated directly from the measured data without use of \eqref{13}. 
Although the existence of \eqref{13} might be guaranteed with the method of inertial manifold \cite{Temam:1997}, its {\color{black}verification and derivation of conditions or assumptions are} 
in our future research. 

\subsection{Derivation of State-Feedback Controller}
\label{subsec:導出}

Our control design is to suppress a target oscillation embedded in \eqref{eqn:KMD} by assigning damping to the associated Koopman mode. 
From \eqref{eqn:KMD0}, the \emph{mode variable} $z_\nu\in\mathbb{C}$ associated with the pure-imaginary eigenvalue $\nu=\ii\omega$ is introduced using the Koopman eigenfunction $\phi_\nu$ as follows:
\begin{equation}
z_{\ii\omega}:=\phi_{\ii\omega}(\vct{x}).
\label{z_nu}
\end{equation}
From \eqref{eqn:KMD}, the absolute value of $z_{\ii\omega}$, namely $|z_{\ii\omega}|=|\phi_{\ii\omega}(\vct{x})|$, governs the magnitude of the sustained single-frequency component. 
To investigate the time evolution of $z_{\ii\omega}$, 
by differentiating $z_{\ii\omega}$ in \eqref{z_nu} with respect to $t$, 
the ODE of $z_{\ii\omega}$ is derived as
\begin{align}
\dot{z}_{\ii\omega} &=\vct{\nabla}\phi_{\ii\omega}(\vct{x})^\top\dot{\vct{x}} \nonumber\\
&= \vct{\nabla}\phi_{\ii\omega}(\vct{x})^\top\{\vct{F}(\vct{x})+\mathsf{B}\vct{u}\} \quad(\because \eqref{13}) \nonumber\\
&= \ii\omega z_{\ii\omega}+\vct{\nabla}\phi_{\ii\omega}(\vct{x})^\top\mathsf{B}\vct{u} \quad(\because \eqref{4}).
\label{a3}
\end{align} 
Under $\vct{u}=\vct{0}$, i.e. under no control, we see
\begin{equation}
\dot{z}_{\ii\omega}=\ii\omega z_{\ii\omega},
\label{a4}
\end{equation}
that is, $|z_{\ii\omega}(t)|$ does not change in $t$, implying that the target component is indeed sustained.  
\emph{Our control design is to assign damping to the system \eqref{a3} so that $|z_{\ii\omega}(t)|$ decreases in $t$}. 
To this end, the so-called polar representation $z_{\ii\omega}=r\ee^{\ii\varphi}$ ($r>0, -\pi\leq\varphi<\pi$) is introduced, 
and then the following ODE of $r$ is derived as
\begin{equation}
2\dot{r}
=\left\{{\rm e}^{-{\rm i}\varphi}\vct{\nabla}\phi_{\ii\omega}(\vct{x})^\top
+{\rm e}^{{\rm i}\varphi}\vct{\nabla}\phi_{-\ii\omega}(\vct{x})^\top\right\}\mathsf{B}\vct{u},
\label{a9}
\end{equation}
where $\overline{\phi_{\ii\omega}}=\phi_{-\ii\omega}$. 
Thus, we design $\vct{u}$ such that $\dot{r}=-Dr$ holds 
where $D>0$ is the design parameter to determine the degree of damping (decaying of amplitude).  
A form of the designed input $\vct{u}$ is the following: 
\begin{align}
\vct{u}=&
-D\mathsf{B}^{+}\left\{\frac{{\rm Re}[\phi_{\ii\omega}(\vct{x})]}{|\phi_{\ii\omega}(\vct{x})|}
\nabla{\rm Re}[\phi_{\ii\omega}(\vct{x})]^\top\right.
\left. +\frac{{\rm Im}[\phi_{\ii\omega}(\vct{x})]}{|\phi_{\ii\omega}(\vct{x})|}
\nabla{\rm Im}[\phi_{\ii\omega}(\vct{x})]^\top\right\}^{+}|\phi_{\ii\omega}(\vct{x})|,
\label{14}
\end{align}
where $\mathsf{B}^{+}$ is the Moore-Penrose pseudo-inverse of $\mathsf{B}$. 
This $\vct{u}$ is a form of state-feedback controllers that is a function of $\vct{x}$. 
Note that the design is closely related to the energy control \cite{Astrom_AUTO36}: see Appendix~\ref{app:remark} as a remark using a simple model of mass point. 
An important point is that the Koopman eigenfunction $\phi_{\ii\omega}$ is estimated as in \eqref{estimated} 
directly from time-series data of the sampled $\vct{x}(t)$ without knowledge of the latent model \eqref{13}. 
In this sense, the designed input $\vct{u}$ is realized in a fully data-driven manner. 


\subsection{Geometric Insight to Performance Evaluation}

We now provide a geometric insight of the control design that is applicable to its performance evaluation. 
The system \eqref{a4} under no control shows that the following subset of the state space $\mathbb{R}^M$, 
\begin{equation}
L_{r}:=\{\vct{x}\in\mathbb{R}^{M}\ |\ |\phi_{\ii\omega}(\vct{x})|=r\},\ \ r>0,
\label{Lc1}
\end{equation}
is invariant under the drift of \eqref{13}. 
The drift (i.e. uncontrolled) trajectory $\vct{x}(t)$ moves \emph{along} the level set $L_{r}$ based on the Koopman eigenfunction $\phi_{\ii\omega}$. 
Also, the level set based on the argument of $\phi_{\ii\omega}$, 
\begin{equation}
L_{\varphi}:=\{\vct{x}\in\mathbb{R}^{M}\ |\ \angle\phi_{\ii\omega}(\vct{x})=\varphi \},\ \ -\pi\leq\varphi<\pi,
\label{Lc2}
\end{equation}
is introduced, and the drift trajectory moves \emph{across} the level set $L_{\varphi}$. 
It should be noted that their observation appears in literature for which the so-called isostable and isochron are developed \cite{Mauroy_CHAOS22,Mauroy_PD261}. 
Our control design is intended such that $r(t)=|z_{\ii\omega}(t)|=|\phi_{\ii\omega}(\vct{x}(t))|$ decreases in $t$. 
Thus, it is expected that the controlled trajectory moves \emph{across} the level sets of $L_{r}$ in a manner that $r(t)$ decreases in $t$. 
The relationship between (un)controlled trajectories and level sets of Koopman eigenfunctions will be numerically investigated in the next section as a performance evaluation of the control design. 


\section{Numerical Simulations}
\label{sec:5}

\subsection{Simulation Setting and Model Validation}

The setting for simulation of the infinite-dimensional model in Sec.\,\ref{subsec:infinite-model} is first summarized. 
We use the Neumann boundary condition for the wall, door, and window in Fig.\,\ref{fig1.5}. 
For instance, on the right and left boundaries in Fig.\,\ref{fig1.5}, the condition is represented as
\begin{equation}
\left.
\begin{aligned}
\frac{\DD}{\DD x}\theta(\vct{r},t) &= K\sub{conv}(\vct{r})\{\theta\sub{ext}(t)-\theta(\vct{r},t)\} \\
\frac{\DD}{\DD y}\theta(\vct{r},t) &=0
\end{aligned}
\right\},
\end{equation}
where $\vct{r}$ is on the boundary and $\theta\sub{ext}(t)$ the outside temperature given as the exogenous single (in which we use the measured data during the same period as in Fig.\,\ref{fig:measurement} and omit it because of the limitation of space). 
The position-dependent parameter $K\sub{conv}(\vct{r})$ is related to convective heat transfer through the boundary. 
It is set at zero for the wall (no heat convection is assumed) and set at the following formula for the window (and door on the upper boundary in Fig.\,\ref{fig1.5}):
\begin{equation}
K\sub{conv}(\vct{r})=\frac{W}{\rho c\sub{p}D\sub{eff}},
\end{equation}
where $W$ is the thermal transmission rate and set at $0.091\,\U{W/(m^2\cdot K)}$. 
Note the the similar condition is used for the upper and lower boundaries in Fig.\,\ref{fig1.5}. 
The other parameters are set as $D\sub{eff}=0.5\,\U{m^2/s}$, $\rho=1.176\,\U{kg/m^3}$, $c\sub{p}=1.007\,\U{J/(kg\cdot K)}$, $T\sub{s}=17\,\U{min}$, $L=6\,\U{min}$, $\theta\sub{ref}=27\,\U{deg.C}$, $V^{\rm on}\sub{air}=0.25\,\U{m^3/s}$, and $V^{\rm off}\sub{air}=0.13\,\U{m^3/s}$. 
The outlet temperature $\theta\sub{air}$ is assumed to be constant at $17\,\U{deg.C}$ in this paper.  
The solution of the linear PDE is derived using the standard center-difference scheme in space (its step sizes $\Delta x=1.8\,\U{m}$ and $\Delta y=1.35\,\U{m}$) and the Euler scheme in time ($\Delta t=1\,\U{s}$).

\begin{figure}[t]
\centering
\includegraphics[width=0.48\textwidth]{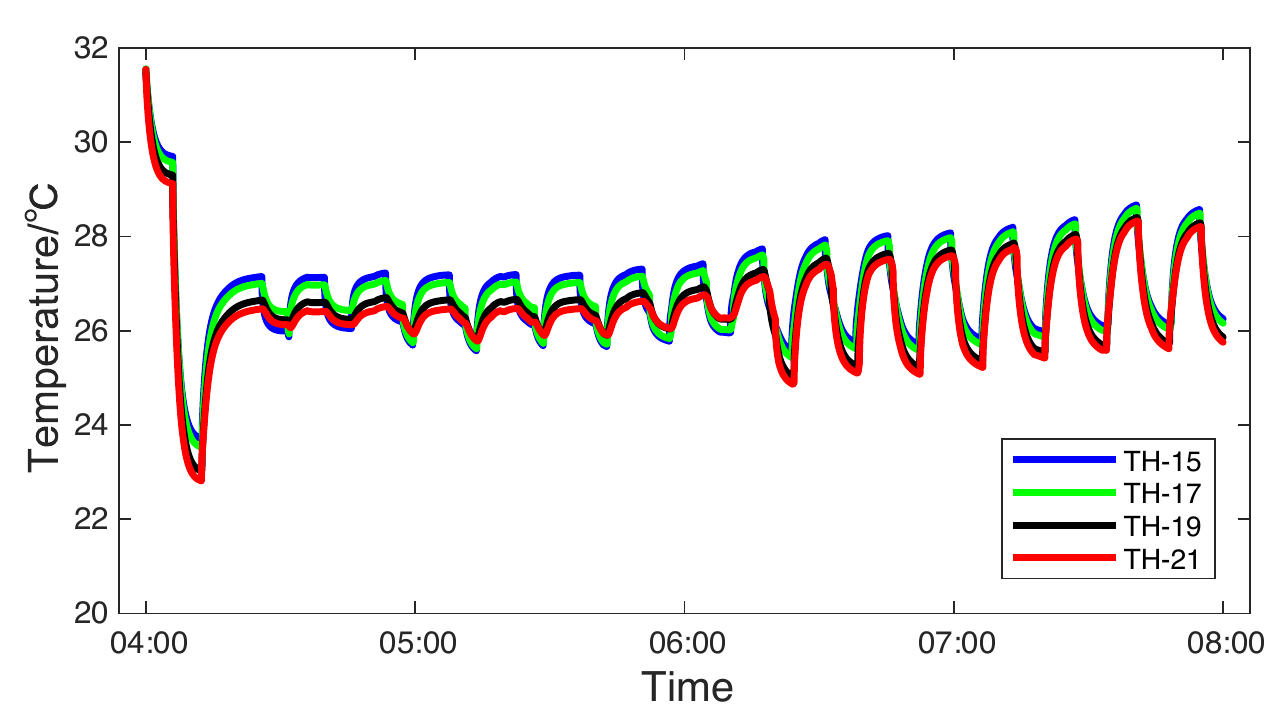}
\caption{%
Open-loop simulation of oscillatory response of in-room temperature field, which is consistent with the measurement data in Fig.\,\ref{fig:measurement}.
}
\label{fig:open-loop}
\end{figure}

Figure~\ref{fig:open-loop} 
shows an open-loop simulation of the infinite-dimensional model in Sec.\,\ref{subsec:infinite-model}, in which the temperature at the four locations close to the four PTACs is plotted.
The setting above is intended for numerically simulating the measurement data in Fig.\,\ref{fig:measurement}. 
It is confirmed in Fig.\,\ref{fig:open-loop} that the temperature oscillates around the set-point, 27\,deg.C. 
The oscillatory patterns in Figs.\,\ref{fig:measurement} and \ref{fig:open-loop} are similar, suggesting that the infinite-dimensional model in Sec.\,\ref{subsec:infinite-model} is validated in terms of the measurement. 

\subsection{Simulation Results}
\label{subsec:results}

\begin{figure*}[t]
\centering
\includegraphics[width=0.95\textwidth]{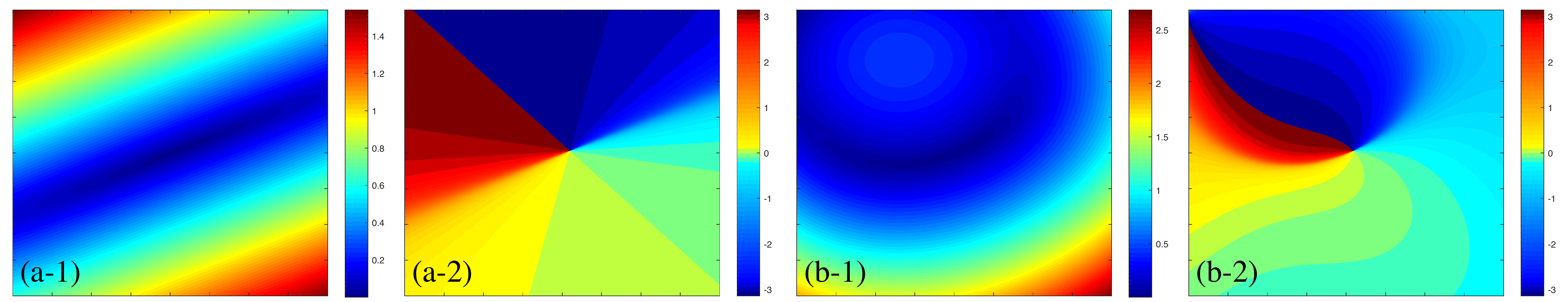}
\caption{Level sets of estimated Koopman eigenfunctions: 
(a-1) absolute value and (a-2) argument for the linear approximation \eqref{eqn:basis}; 
(b-1) absolute value and (b-2) argument for the nonlinear approximation \eqref{eqn:basis}. 
The range of the visualization is $(x_1,x_3)\in[-2,2]\times[-2,2]$ at $x_2=x_4=0$.}
\label{fig:KEF}
\end{figure*}

\begin{figure*}
\centering
\begin{minipage}[h]{\textwidth}
\centering
\includegraphics[width=0.48\textwidth]{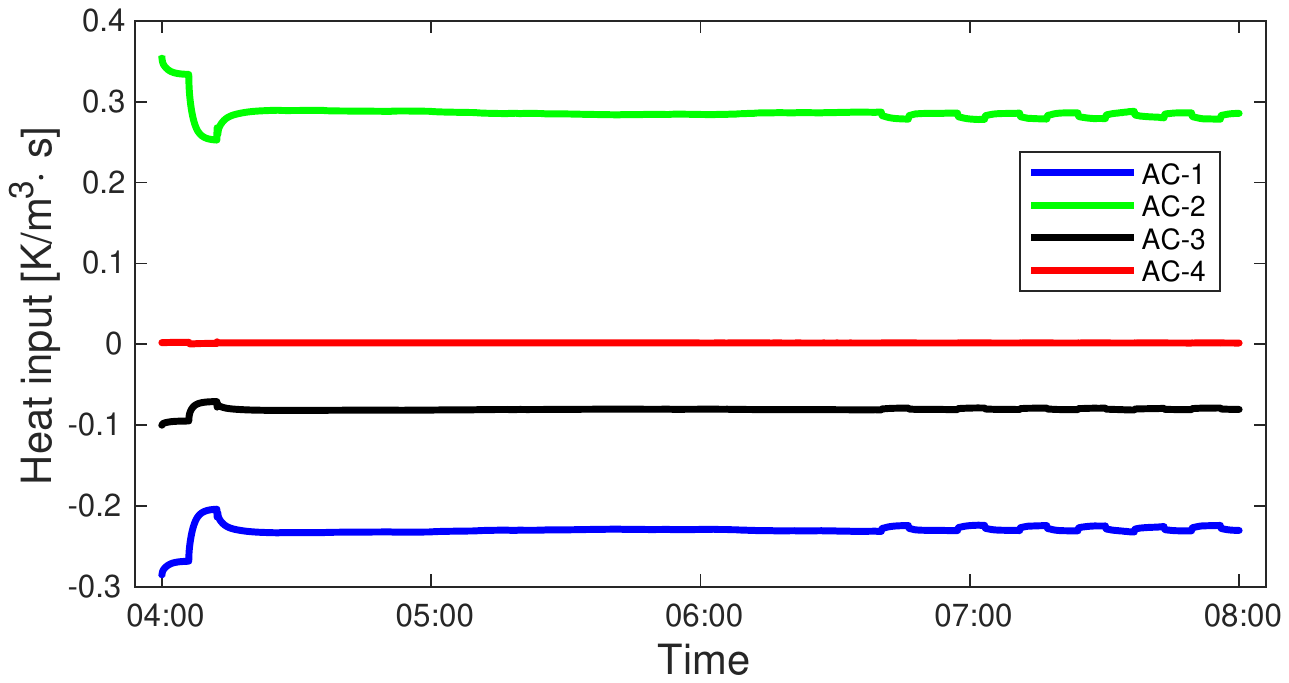}
\includegraphics[width=0.48\textwidth]{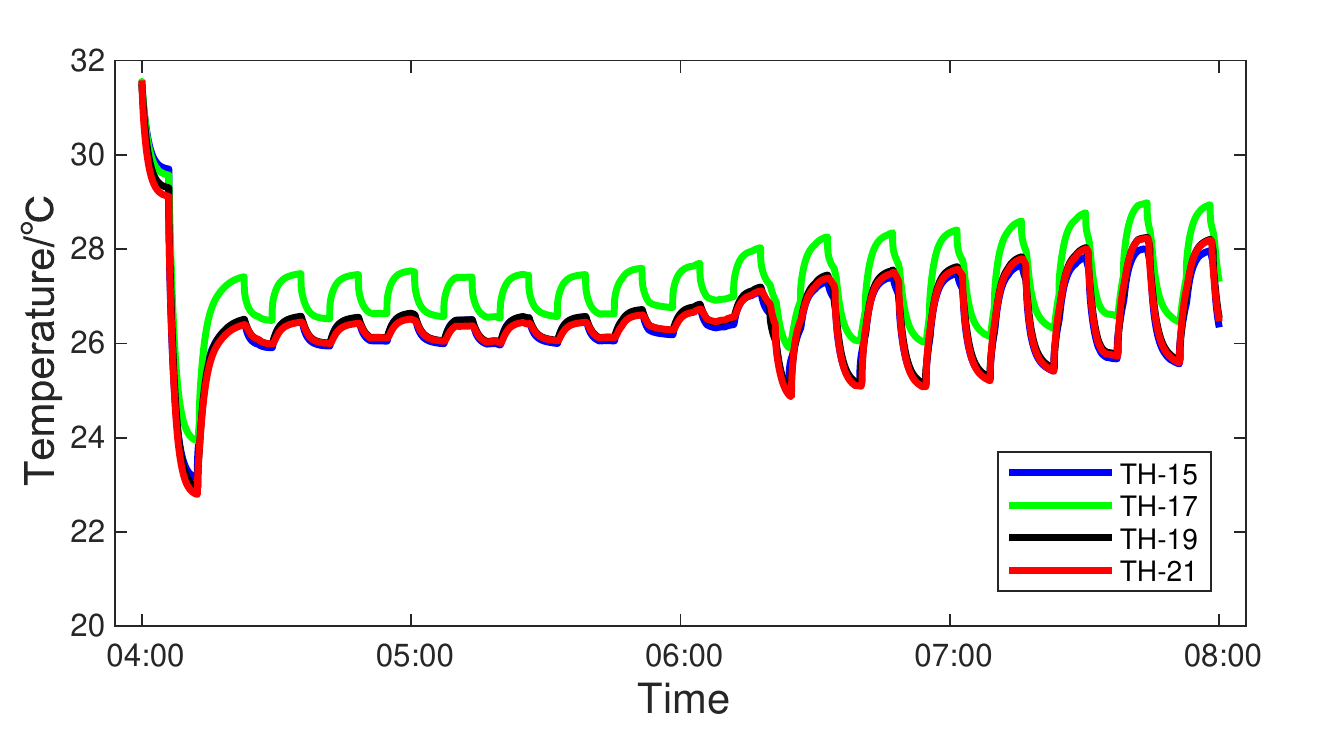}
\vspace*{-2mm}
\subcaption{Use of the linear approximation \eqref{eqn:basis} in Fig.\,\ref{fig:KEF}a}
\label{fig:制御後H}
\end{minipage}
\begin{minipage}[h]{\textwidth}
\centering
\includegraphics[width=0.48\textwidth]{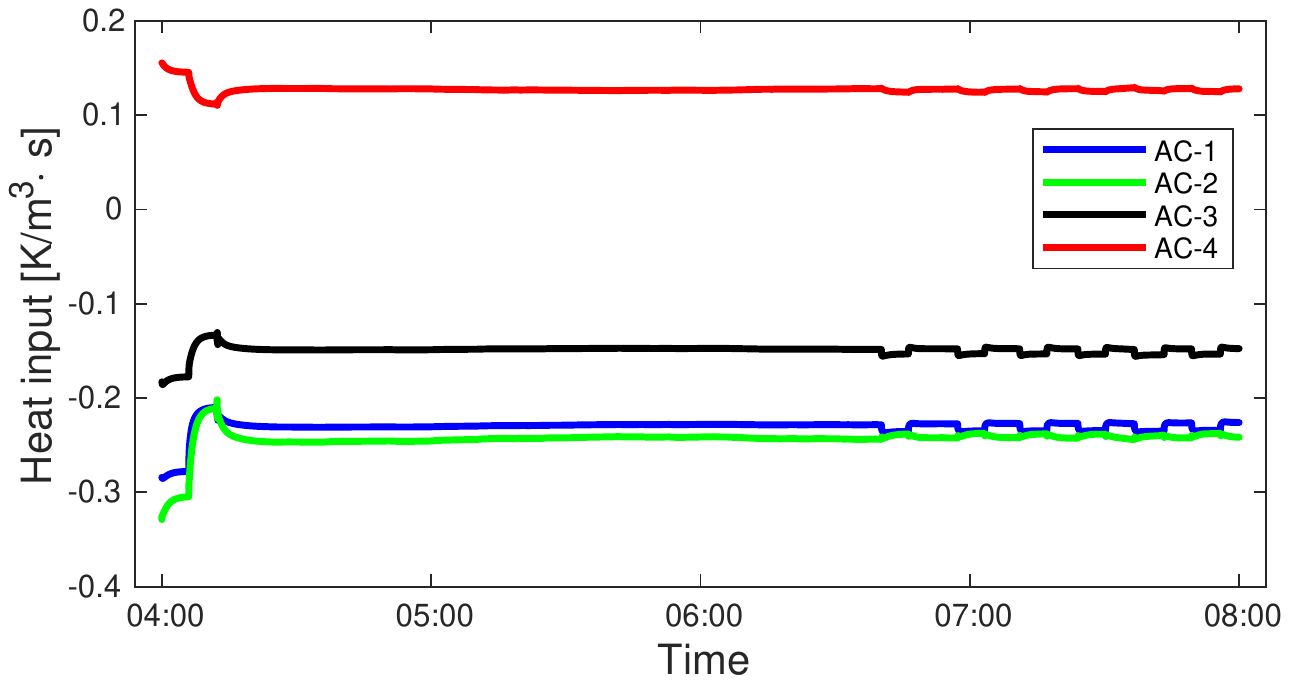}
\includegraphics[width=0.48\textwidth]{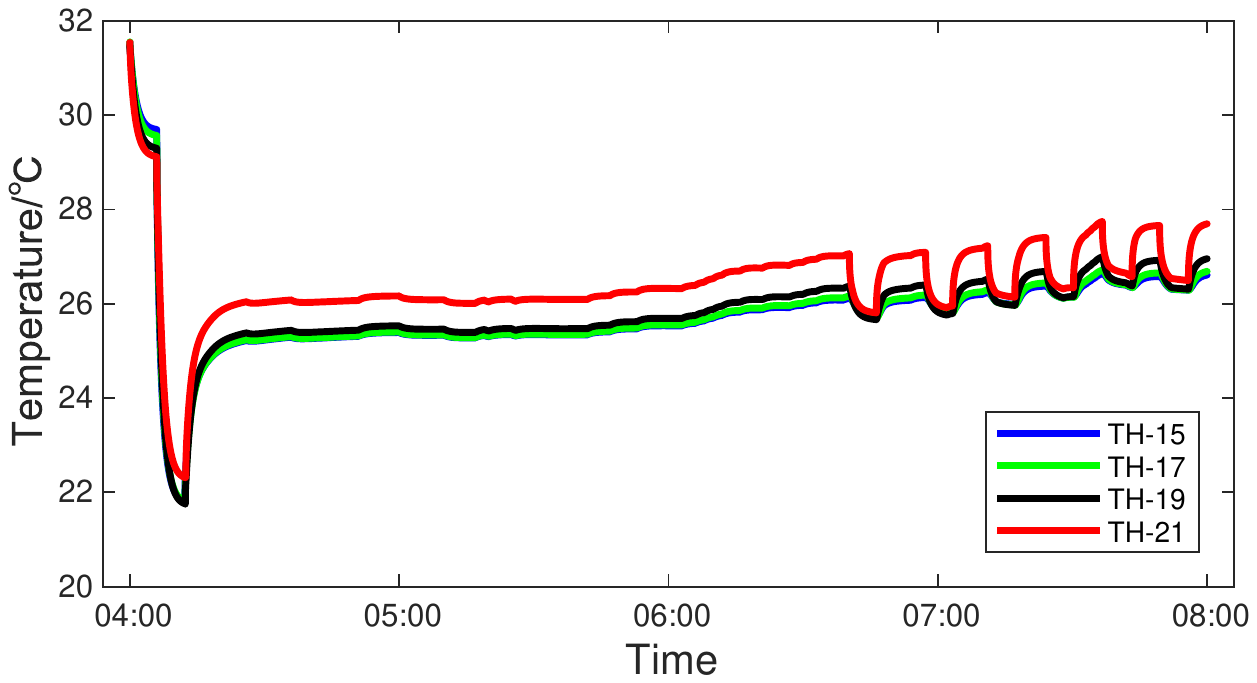}
\vspace*{-2mm}
\subcaption{Use of the nonlinear approximation \eqref{eqn:basis} in Fig.\,\ref{fig:KEF}b}
\label{fig:制御後E}
\end{minipage}
\caption{Closed-loop simulations of oscillatory responses of in-room temperature field, which are produced with the infinite-dimensional system in Sec.\,\ref{subsec:infinite-model} with the designed input}
\label{fig:closed-loop}
\end{figure*}

We consider closed-loop simulations of the in-room temperature field. 
For this, the EDMD is conducted for estimating Koopman eigenvalues and eigenfunctions from the time-series data in Fig.\,\ref{fig:open-loop}.  
In this paper, we choose two different sets of basis functions $\vct{\gamma}(\vct{x})$ as
\begin{equation} 
\vct{\gamma}(\vct{x})=
\left\{
\begin{array}{ll}
[1\ \vct{x}^\top]^\top & \textrm{(linear)}, \\\noalign{\vskip 1mm}
[1\ \vct{x}^\top\ (\vct{x}^2)^\top\ (\vct{x}^3)^\top]^\top 
& \textrm{(nonlinear)},
\end{array}
\right.
\label{eqn:basis}
\end{equation}
where $\vct{x}^{n}:=[x^n_1,\ x^n_2,\ x^n_3,\ x^n_4]$. 
By EDMD with the nonlinear approximation, we found an almost sustained Koopman mode with period $13.98\,\U{min}$. 
The period is close to that in the measurement data in Fig.\,\ref{fig:measurement}, for which we have an almost sustained mode with period $15.31\,\U{min}$. 
The level sets of the estimated Koopman eigenfunctions are shown in Fig.\,\ref{fig:KEF} which will be used in Sec.\,\ref{subsec:PE}. 
The shapes of the level sets (visualized in common color) are clearly affected by the presence of nonlinearity in the basis function $\vct{\gamma}(\vct{x})$. 
Figure~\ref{fig:closed-loop} shows closed-loop simulations of oscillatory responses of the in-room temperature field. 
The two left figures are the heat inputs generated by \eqref{14}, and the two right figures are the temperature at the four locations \textsf{TH-15}, \textsf{TH-17}, \textsf{TH-19}, and \textsf{TH-21}. 
The figure (a) is the result based on the linear approximation \eqref{eqn:basis}, and the figure (b) based on the nonlinear approximation \eqref{eqn:basis}. 
To make control efforts equal in the linear and nonlinear cases for fair comparison, the energy norm of the heat input $u_1$ at \textsf{AC-1} (\emph
{blue} in the left figures), $\left(\int_{0}^{T}u_{1}(t)^{2}\dd t\right)^{1/2}$ with  $T=14400\,\U{s}$, is the same in Figs.\,\ref{fig:制御後H} and \ref{fig:制御後E} by adjusting the design parameter $D$ ($0.224$ for the linear and $0.035$ for the nonlinear). 
It is shown in the figures that the input based on the nonlinear approximation effectively works for the suppression than the linear approximation. 
In Fig.\,\ref{fig:制御後H}, the heat input $u_2$ (\emph{green}) at \textsf{AC-2} is positive and $u_4$ (\emph{red}) at \textsf{AC-4} is positive and close to zero, while in Fig.\,\ref{fig:制御後E}, $u_2$ (\emph{green}) is negative and $u_4$ (\emph{red}) is positive. 
These heat inputs are adequately designed in the nonlinear approximation for the effective assignment of damping. 
Note that the oscillations after around 6:30am are not sufficiently suppressed in both the approximations. 
This might be because for this duration, another Koopman mode is excited as indicated in Fig.\,\ref{fig:open-loop} and discussed below.


\subsection{Performance Evaluation}
\label{subsec:PE}

\begin{figure*}[t]
\centering
\begin{minipage}[h]{\textwidth}
\centering
\includegraphics[width=0.495\textwidth]{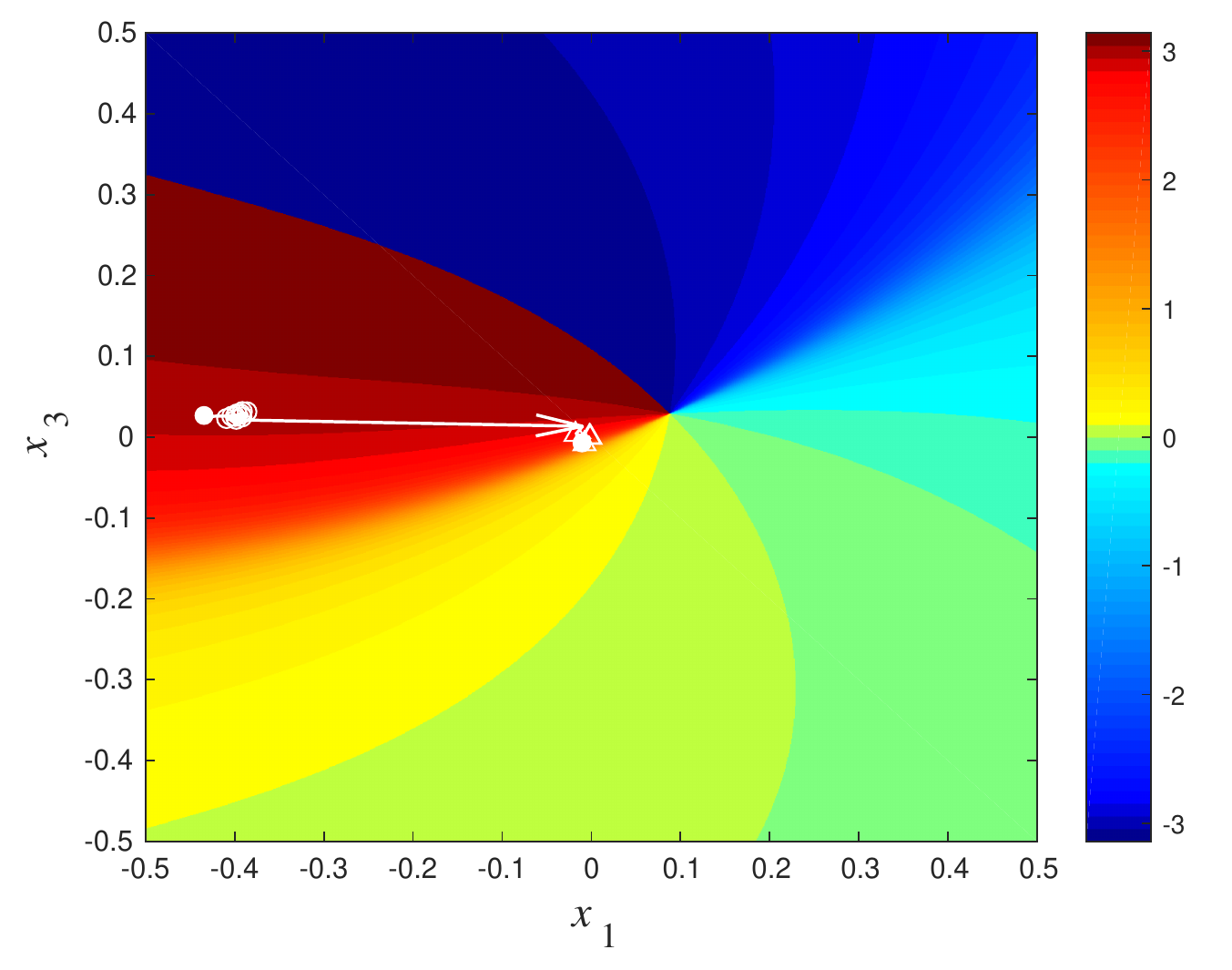} 
\includegraphics[width=0.495\textwidth]{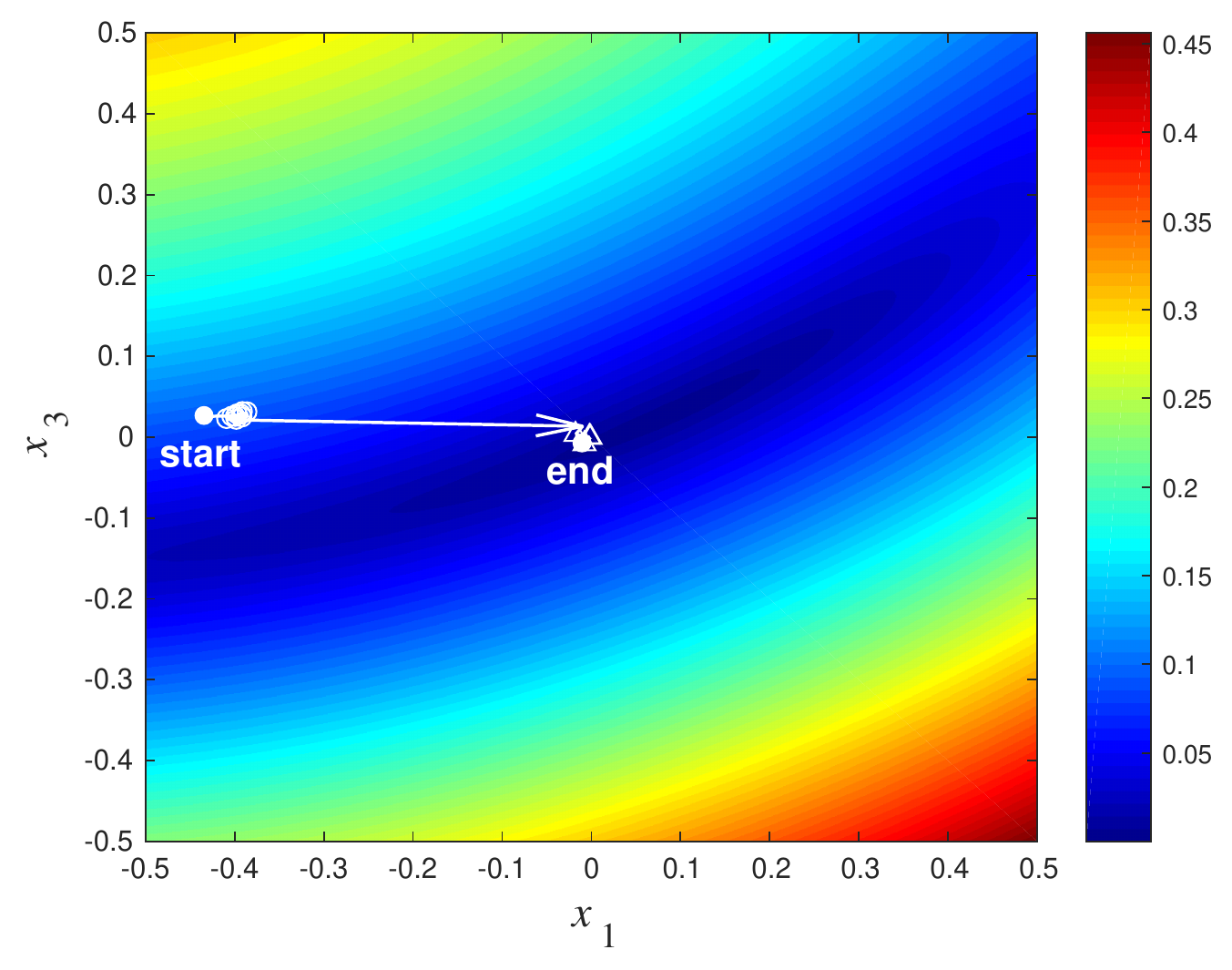}
\vskip -5mm
\subcaption{Without control}
\label{fig:制御前_分類}
\end{minipage}
\begin{minipage}[h]{\textwidth}
\centering
\includegraphics[width=0.495\textwidth]{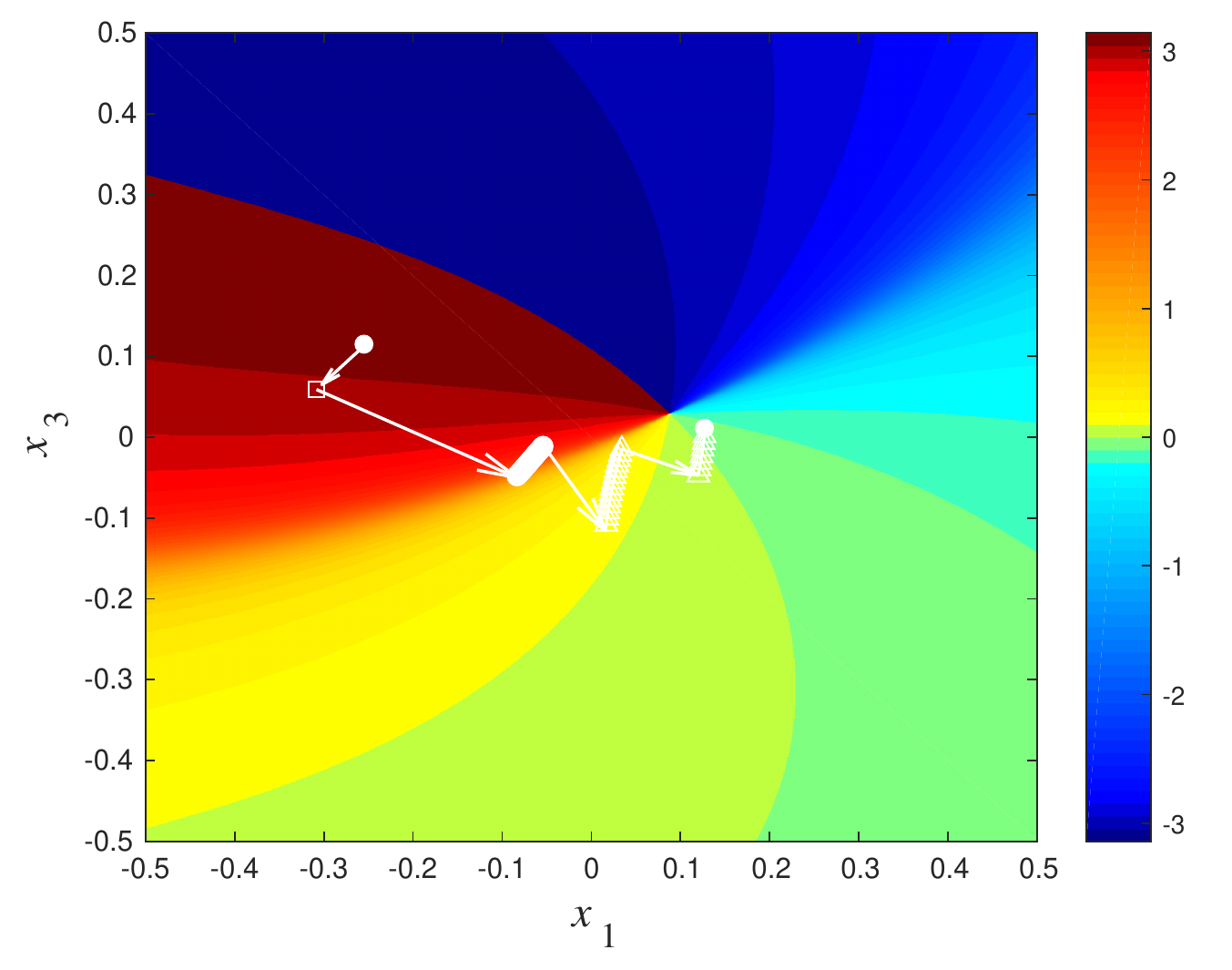}
\includegraphics[width=0.495\textwidth]{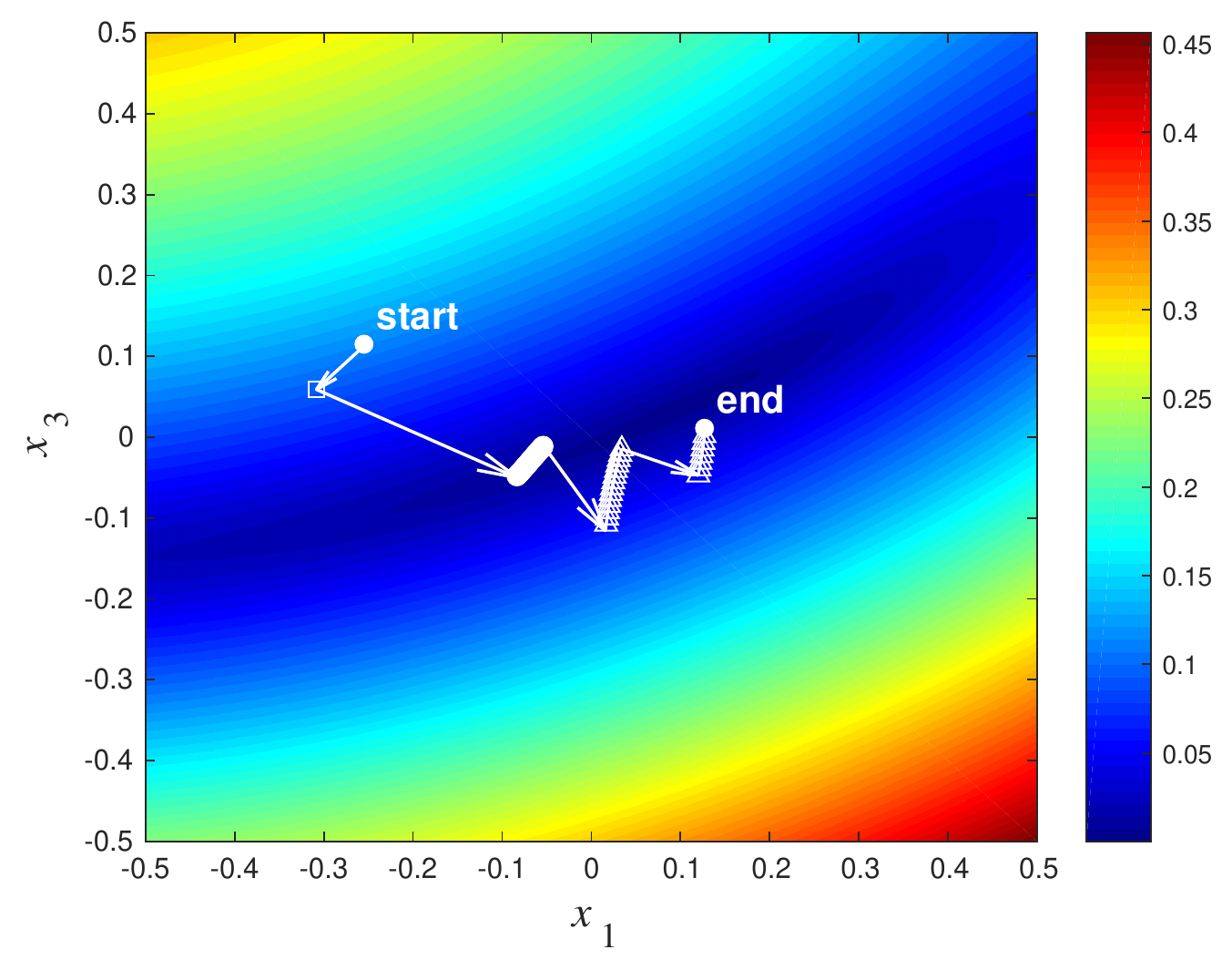}
\vskip -5mm
\subcaption{With control}
\label{fig:制御後_分類}
\end{minipage}
\caption{(Un)controlled trajectories and their projection onto level sets of the estimated Koopman eigenfunction in Fig.\,\ref{fig:KEF}b: 
(left) absolute value and (right) argument. 
The range of the visualization is $(x_1,x_3)\in[-0.5,0.5]\times[-0.5,0.5]$ at $x_2=x_4=0$.}
\label{fig:分類}
\end{figure*}

Lastly, we evaluate the control performance in terms of the level sets of a Koopman eigenfunction. 
In particular, the Koopman eigenfunction estimated via the nonlinear approximation in Fig.\,\ref{fig:KEF}b is used because this case shows a better performance in Fig.\,\ref{fig:closed-loop}. 

Figure~\ref{fig:分類} shows (un)controlled trajectories and their projection onto level sets of the estimated Koopman eigenfunction in Fig.\,\ref{fig:KEF}b. 
The left figures are for the argument $L_\varphi$ and the right figures are for the absolute value $L_r$. 
The values of $\varphi$ and $r$ are depicted in the color bars. 
The visualization of the level sets is for the range $(x_1,x_3)\in[-0.5,0.5]\times[-0.5,0.5]$ at \emph{fixed} $x_2=x_4=0$. 
To do the projection, we set a small positive value $\epsilon$ and pick up a point from the trajectory $\vct{x}(t)=[x_1(t),x_2(t),x_3(t),x_4(t)]^\top$ of the temperature at the four locations \textsf{TH-15}, \textsf{TH-17}, \textsf{TH-19}, and \textsf{TH-21}, which satisfy $|x_2(t)|,\ |x_4(t)|<\epsilon$: this is similar to the construction of Poincar\'e map.  
The start and end points are indicated by \emph{white} dots. 
The square ($\Box$) represents the picked states during the period from 4:00am to 4:30am, the circle ($\bigcirc$) does during the period from 4:30pm to 6:30pm, and the triangle ($\triangle$) does during the period from 6:30am to 8:00am. 
The uncontrolled case in Fig.\,\ref{fig:制御前_分類} shows that there are many circles ($\bigcirc$) kept near the start point and many triangles ($\triangle$) kept near the end point. 
This indicates that the absolute value of the Koopman eigenfunction along the uncontrolled trajectory does not change during the period from 4:00am to 6:30am, which validates the invariance of the level sets of the Koopman eigenfunction without control. 
The transition between $\bigcirc$ and $\triangle$ might be due to the excitation of another Koopman mode indicated in the open-loop simulation of Fig.\,\ref{fig:open-loop}. 
The controlled case in Fig. \ref{fig:制御後_分類} shows that the change of absolute value of the Koopman eigenfunction occurs from the start point to $\bigcirc$ in a descent manner, which is consistent in our control design. 
The projections in Fig.\,\ref{fig:closed-loop} show that the heat input \eqref{14} works as designed in the geometric perspective.

\section{Conclusions}
\label{sec:6}

This paper demonstrated the data-driven damping assignment to Koopman mode by its application to the design of controller for suppressing an oscillatory response of temperature field in a building. 
We presented numerical simulations of thermal dynamics guided by measurement of a practical room space, thereby showing the effectiveness of the damping assignment. 

Our control design exploits spectral properties of the Koopman operator for the drift, namely open-loop dynamics. 
This is similar to literature in the Koopman operator framework, see, e.g., \cite{Sootla_AUTO91} in which the authors exploit isostables for optimal control. 
In comparison with a large number of literature in applications of Koopman-based lifting to control, we contend that our control design utilizes the geometric perspective of nonlinear systems extracted via the spectral properties. 

Several studies following this research are possible. 
{\color{black}First, in} 
the application viewpoint, it is necessary to verify our control design in near practical situations with three dimension. 
This is important in terms of its robust performance and will be reported in another opportunity.  
{\color{black}Second, it} 
is 
important to clarify how the idea is implemented in practical air-conditioning systems. 
{\color{black}Regarding this,  a comparative study in terms of classical control design is required. 
Also, it is interesting to adjust the dimensions of the control and state ($m$ and $M$) for better control performance. 
Lastly, in} 
the theoretical viewpoint, we investigated how the dynamics described by the nonlinear PDE were actuated on finite points on its physical domain. 
Its theoretical analysis for guaranteeing performance and limitation is interesting. 
Regarding this, it is necessary to consider the problem of decoupling for damping assignment to multiple Koopman modes, although it was not needed for  consideration in this paper. 


\appendix
\section{Extended Dynamic Mode Decomposition}
\label{app:EDMD}

In this appendix, we summarize the algorithm called the Extended Dynamic Mode Decomposition (EDMD) \cite{Matt_JNLS25,Klus_JCD3} in order to estimate Koopman eigenvalues and eigenfunctions \eqref{3} directly from time series data. 
For this, consider the state's trajectory $\vct{x}(t)$ of \eqref{1} and its equally-spaced sampling with period $h$, denoted by $\vct{x}_k:=\vct{x}(kh)$ for $k=0,\ldots,m$ ($m$ is a natural number). 
The following snapshot matrices are then considered:
\begin{equation}
\left.
\begin{split}
\mathsf{X}&:=[\vct{x}_0,\ldots,\vct{x}_{m-1}]\in\mathbb{R}^{n\times m}\\
\mathsf{X}^{\prime}&:=[\vct{x}_1,\ldots,\vct{x}_m]\in\mathbb{R}^{n\times m}
\end{split}
\right\}.
\label{5}
\end{equation}
Also, as introduced in Sec.\,\ref{sec:3}, a set of user-specified basis functions (observables) with $q$ number,  $\vct{\gamma}(\vct{x}):=[\gamma_{1}(\vct{x}),\ldots,\gamma_{q}(\vct{x})]^{\top}$: $\mathbb{R}^{n}\rightarrow\mathbb{R}^{q}\ 
(\gamma_{j}\in\mathcal{K})$ are chosen. 
By using this $\vct{\gamma}$ and \eqref{5}, the following matrices are computed:
\begin{equation}
\left.
\begin{split}
\mathsf{\Gamma}_{\mathsf{X}}&:=[\vct{\gamma}(\vct{x}_0),\ldots,\vct{\gamma}(\vct{x}_{m-1})] \in\mathbb{R}^{q\times m}\\
\mathsf{\Gamma}_{\mathsf{X}^{\prime}}&:=[\vct{\gamma}(\vct{x}_1),\ldots,\vct{\gamma}(\vct{x}_{m})] \in\mathbb{R}^{q\times m}
\end{split}
\right\}.
\label{7}
\end{equation}
Then, a matrix representation $\mathsf{U}\in\mathbb{R}^{q\times q}$ of the Koopman operator ${U}^{h}$ is introduced as follows:
\begin{equation}
\mathsf{U}:=\mathsf{\Gamma}_{\mathsf{X}^{\prime}}\left(\mathsf{\Gamma}_{\mathsf{X}}\right)^{+}.
\label{8}
\end{equation}
The eigenvalues $\lambda_j$ ($j=1,\ldots,q$) of $\mathsf{U}$ are approximations of eigenvalues of ${U}^{h}$ (in the discrete-time sense). 
The Koopman eigenvalue $\nu_j$ in this paper is determined through conversion,
\begin{equation}
\nu_j=\frac{\ln\lambda_j}{h}.
\end{equation}
The left-eigenvectors $\vct{\xi}_j$ of $\mathsf{U}$, denoted by $\mathsf{\Xi}:=[\vct{\xi}_1,\ldots,\vct{\xi}_q]^\top$, provide approximations of the Koopman eigenfunctions in \eqref{estimated}:
\begin{equation}
\vct{\phi}(\vct{x}_k)=\mathsf{\Xi}\vct{\gamma}(\vct{x}_k),
\label{9}
\end{equation}
where they are valid on the sampled trajectory and are used in this paper for a subspace of the state space.

\section{Remark on the Proposed Damping Assignment}
\label{app:remark}

Our control design is closely related to the energy control \cite{Astrom_AUTO36}. 
As a simple example, let us consider the equation of motion for a mass point connected to a linear spring with an external force, given by
\begin{equation}
m\frac{\dd^2x}{\dd t^2}=-kx+u,
\end{equation}
where $x$ is the one-dimensional displacement, $m$ the mass constant, $k$ the spring coefficient, and $u$ the external force.  
By defining the momentum $p:=m\dot{x}$, the following linear dynamical system is derived as
\begin{equation}
\left[
\begin{array}{c}
\dot{x} \\ \dot{p}
\end{array}
\right]
=
\left[
\begin{array}{cc}
0 & 1/m \\ -k & 0
\end{array}
\right]
\left[
\begin{array}{c}
x \\ p
\end{array}
\right]
+
\left[
\begin{array}{c}
0 \\ 1
\end{array}
\right]
u.
\label{eqn:mass-pnt}
\end{equation}
Here, it is possible to show that the energy of the mass point,
\begin{equation}
\phi_0(x,p):=\frac{kx^2}{2}+\frac{p^2}{2m},
\end{equation}
is a Koopman eigenfunction associated with eigenvalue $0$ for the system \eqref{eqn:mass-pnt} with $u=0$ (this can be checked using \eqref{4}):
\begin{equation}
{U}^t\phi_0(x,p)=\ee^{0t}\phi_0(x,p).
\end{equation}
The level sets of $\phi_0$ are level sets of energy of the mass point and clearly invariant under the uncontrolled motion ($u=0$).  
For our design, the ODE of the modal variable $r=\phi_0(x,p)$ is given by\footnote{Unlike in the main body, because the Koopman eigenfunction and eigenvalue are real-valued, the mode variable in this example becomes real-valued.}{\color{black}
\begin{equation}
\dot{r}=\frac{p}{m}u.
\end{equation}
Thus, for a target level $\bar{r}$ of the energy, 
$u$ is designed so that $\dot{r}=-D(r-\bar{r})$ (where $D>0$) holds:
\begin{equation}
u=-\frac{mD}{p}(r-\bar{r})
=-\frac{mD}{p}\left(\frac{kx^2}{2}+\frac{p^2}{2m}-\bar{r}\right), 
\end{equation}
where $p=0$ should be taken care by another idea.  
This design is intended for controlling} 
the energy $r$ of the mass point by assigning damping to the Koopman mode with eigenvalue $0$. 



\end{document}